\documentclass[pra,twocolumn,superscriptaddress,showpacs,amsmath,amssymb]{revtex4-2}

\usepackage{graphicx}% Include figure files
\usepackage{dcolumn}% Align table columns on decimal point
\usepackage{bm}% bold math
\usepackage{xcolor}
\usepackage{float}
\usepackage{amsmath}

\begin{document}

\title{Quantum determinism and completeness restored by indistinguishability and long-time particle detection}

\author{Patrick Navez}
\affiliation{Laboratoire Charles Coulomb UMR 5221 CNRS-Universit\'e de Montpellier, F-34095 Montpellier, France}
\author{Henni Ouerdane}
\affiliation{Skolkovo Institute of Science and Technology, 30 Bolshoy Boulevard, bld. 1, 121205 Moscow, Russia}
\date{\today}

\begin{abstract}
We argue that measurement data in quantum physics can be rigorously interpreted only as a result of a statistical, macroscopic process, taking into account the indistinguishable character of identical particles. Quantum determinism is in principle possible on the condition that a fully-fledged quantum-field-theoretic model is used to describe the measurement device in interaction with the studied object as one system. In contrast, any approach that relies on Born's rule discriminates the dynamics of a quantum system from that of the detector with which it interacts during measurement. In this work, we critically analyze the validity of this measurement postulate applied to single-event signals. In fact, the concept of ``individual'' particle becomes inadequate once both indistinguishability and a scattering approach allowing an unlimited interaction time for an effective detection, are considered as they should be, hence preventing the separability of two successive measurement events. In this context, measurement data should therefore be understood only as a result of statistics over many events. Accounting for the intrinsic noise of the sources and the detectors, we also show with the illustrative cases of the Schr\"odinger cat and the Bell experiment that once the Born rule is abandoned on the level of a single particle, realism, locality and causality are restored. We conclude that indiscernibility and long-time detection process make quantum physics not fundamentally probabilistic.
\end{abstract}

\maketitle

\section{Introduction}
Quantum mechanics is in its original formulation compatible with any experimental result obtained using a statistics-based approach for the measurement of a quantum system's properties. However, the information that can be extracted from measurement data being fundamentally limited and prone to interpretation biases, complete information on a system's quantum state remains inaccessible to human knowledge. Besides complementarity and the Heisenberg uncertainty relations, which impose absolute bounds on the precision of the simultaneous measurement of pairs of observables, the main limiting factor is that no quantum state is immune to a measurement process, and that the obtained data rather reflects the state of the system in interaction with the measurement device. This fact has been extensively discussed in relation with the no-cloning theorem \cite{WoottersZurek1982,Dieks1982}.

To overcome the difficulty of giving an interpretation of the measurement data at the level of elementary particles without an exact, detailed knowledge of the experimental conditions on the source and the detector (thus involving a macroscopic number of particles), full determinism, which is seen as a cumbersome obstacle, has been abandoned in favor of probabilistic predictability with a set of postulates, including the Born rule \cite{Born1926}. One major consequence is that in this framework, properties of quantum systems are assumed to acquire a definite value only because of measurement, thereby implying that there cannot be an ontological continuity between classical and quantum physics: quantum properties become physical quantities only when they are measured. The uncertainty principle makes the matter even more intricate as the precise measurement of one quantum variable (e.g., momentum) renders impossible for its conjugate quantity (position) the acquisition of a definite value. But since the postulates provide a formal ground (based on axioms formulated by Dirac \cite{Dirac1930} and von Neumann \cite{vonNeumann1932}) for both the description of a quantum system and the measurement of its properties in terms of states, observables and their time-evolution, quantum mechanics is commonly taken as an operatorial pragmatic approach based on a set of postulates for measurement and control. An essential question remains though: whether physics at the quantum level fundamentally is probabilistic or deterministic.

Here, we argue that the Born rule (or probability measurement postulate) is a non-essential mathematical convenience that should no longer be used, even though for nearly 100 years, much research in quantum physics has relied upon it for the interpretation of quantum experimental data. In a nutshell, this rule states that the probability density of finding a system in a given state, \emph{immediately after measurement}, is proportional to the square of the amplitude of the system's wave function in that state \emph{immediately before measurement}, thus assuming that the system was isolated from its environment during the measurement process. In this framework, the measurement of a physical quantity can only yield one of the eigenvalues of the corresponding observable with a certain probability. As such, the Born rule separates the dynamics of the studied quantum system from that of the measurement device, thus assuming that measurement data correspond to observables as properties of the studied quantum system only. However, the elephant in the room that has been systematically overlooked in practice, is that there is no reason why quantum mechanics should discriminate between the object of study and the measurement device it is in interaction with as though they were independent entities. A precise knowledge of the interaction between the detector and the quantum system is required for a detailed and rigorous description of the measurement process and hence a sound interpretation of the data. This is a considerable challenge. 

The separation between the quantum or microscopic world and the macroscopic one that includes measurement devices stems from quantum indeterminacy along with the so-called ``Copenhagen interpretation'' (regardless of its many ``flavors'' as Mermin put it \cite{Mermin2017}). More precisely, in this interpretation, separation appears to be necessary because while the wave function of an isolated of quantum system can be defined as a linear superposition of eigenfunctions that satisfy the Schr\"{o}dinger equation, there is no such linear combination that may apply for macroscopic systems (measurement device) states as the classical equations of motion such as, e.g., the Hamilton-Jacobi or Lagrange, are nonlinear. Hence the pervasive concept of probability in the description of quantum phenomena as indeterministic in nature, while probabilities do not appear anywhere during the quantum dynamics but only at the measurement instance, the exact time and the duration of which cannot be rigorously known. 
Consequently, in this paradigm, an observation or measurement of a quantum system's property is an irreversible stochastic process that leads to the collapse of the wave function into one of the isolated (before measurement) system's eigenstates. The resulting wave function after measurement is interpreted as a quantum back action \cite{Devoret2013} and is subsequently governed by the unitary quantum dynamics.

It is important to point out here that, as noted in Ref. \cite{Weschler2021}, there is no certain knowledge about the destruction or preservation of the other eigenstates of the state space after the collapse. None of the foundational works of Born \cite{Born1926}, Dirac \cite{Dirac1930}, von Neumann \cite{vonNeumann1932} and, later, L\"uders \cite{Luders1950} provide a formal proof of the wave function collapse, so one should simply accept that, as already pointed above, properties of quantum systems only ``secure'' their actual value upon measurement, and by that means become physical quantities. Several notable works however, have questioned the Copenhagen interpretation and the wave function collapse, and have provided their own explanation. These include the de Broglie-Bohm theory \cite{deBroglie,Bohm1952I,Bohm1952II}, Everett's ``many-worlds'' interpretation \cite{Everett1957}, the transactional interpretation \cite{Cramer1986}, 't Hooft's cellular automaton interpretation \cite{tHooft2016},  and several others discussed in, e.g., Refs.~\cite{Weschler2021,Bassi2013}. Here, for brevity, we simply note that the de Broglie–Bohm theory, which is non-local yet deterministic, adds a layer of complexity by introducing the pilot equation  and  becomes difficult to extend in a relativistic formulation. And, while the many-worlds interpretation does without Born's rule, it ``splits'' the universe into mutually unobservable alternate histories, which can hardly be proven  experimentally.

As stated in Ref.~\cite{Bassi2013}, either of the following three courses of action can in principle be chosen to manage the conceptual difficulties that plague the interpretation of actual measurement data: i/ a change of interpretation of quantum mechanics; ii/ a change of its mathematical formulation; iii/ a modification of the theory. As of yet, no breakthrough is in sight, and the application of the Born rule that fostered the probabilistic interpretation of measurement outcomes remains prevalent. While this facilitates the use of classical concepts and language to describe quantum phenomena and properties, the trade-off is in the necessity to consider the measurement device as classical \cite{tHooft2016}. Ceasing to apply the Born rule would allow to discard the notion of probabilities being a constitutive element of quantum mechanics thereby paving the way to its deterministic interpretation. The cost would then be a more elaborate description of the detection process. Given the widespread use of the theory in its probabilistic interpretation and its achievements, one may argue that while not minor, the matter remains a specialized problem with no strong consequence for experiments as discussed in a recent editorial~\cite{editorial2022}. However, Werner Heisenberg's assertion that: ``\emph{What we observe is not nature in itself but nature exposed to our method of questioning}'' \cite{Heisenberg}, can serve as a warning against the systematic use of any methodology that eludes fundamental difficulties to interpret experimental data. 

The present work is devoted to the analysis of measurement in quantum physics experiments to provide a more adequate view of the theory by avoiding its probabilistic interpretation. The article is organized as follows. In Section II, we critically discuss the Born rule as a corner stone of most of the research in quantum physics. We show that not only it is possible to dismiss it but that it must be abandoned for the modeling of single-event measurements. In Section III, we discuss key aspects of quantum physics experiments, notably that they are fundamentally based on statistical processes. To substantiate our critical discussion, we provide two illustrative examples: The Schr\"odinger cat experiment and the Bell experiment in Sections IV and V respectively. The article ends with a discussion and concluding remarks.

\section{The inadequacy of the Born rule for single events}
\subsection{Obstacles on the path towards the second revolution}
At the time of the formulation of the Born rule in 1926 \cite{Born1926}, physicists and chemists ambitioned and managed to gain a detailed knowledge and understanding of the electronic structures of atoms and molecules, neglecting the effects of the apparatus upon measurement of a system's quantum properties. Contributions of other authors, including von Neumann \cite{vonNeumann1932}, Dirac \cite{Dirac1927}, and Jordan \cite{Jordan1927a,Jordan1927b}, which adhere to the probabilistic interpretation of quantum mechanics, made the Born rule a postulate of quantum mechanics \cite{proof}. These works were key to move from the early or ``old'' quantum mechanics to its modern formulation; but to foster the development of quantum technology, a second  revolution of quantum mechanics \cite{Dowling,Deutsch2020} has taken place with new theoretical and experimental methods to design and manufacture the next generation of useful quantum devices and processes for, e.g., quantum computing, quantum cryptography, quantum teleportation, and precision metrology. These methods aim to allow manipulation and control of single atoms, photons and ions \cite{Haroche2013,Wineland2013}. As far as measurements are concerned, a huge host of modern literature is devoted to the field, focusing in particular on quantum trajectories, continuous measurements, weak or generalized (ancilla-based) measurements, weak values, and the effect on environment on measurements in open systems \cite{Wiseman,Peres1993,Brun2002,Jacobs2006,Tamir2013,Svensson2013,Jacobs2014,Dressel2014,Jordan2024}. We highlight here that none of these representative works related to measurement are based on a quantum field theory, thus ignoring the formalism of the Fock space for the interpretation of measurement outcomes in the quantum state space fully accounting for quantum statistics and long-time detection.

Further, the joint system-detector evolution and the corresponding back-action effects are nowadays one of the cornerstones of the measurement-based quantum state engineering, performed experimentally on various quantum-hardware platforms \cite{Devoret2013}. Even so, these developments are based on rules akin to the Born rule (L\"uders' and von Neumann's projection postulates, and stochastic approach) \cite{Gisin1984,Patel2017}, lacking the heuristic approach necessary to grasp the whole physics aspects associated to a physical phenomenon.
It misses the fundamental question as to why quantum randomness should be considered as an irreducible feature of quantum physics. Without a measurement procedure necessary for quantum engineering and made by humans, any quantum unitary evolution of an isolated system does not made any use of randomness. It is therefore the simplifying assumptions (trace of the noise bath and detector description) that renders unavoidable the use of probabilities. 

In light of the above, the two assertions related to randomness in Ref.~\cite{Dowling}:\\ 

``\emph{Uncertainty principle: for every perfectly specified quantum state there is always at least one measurement, the results of which are completely certain, and simultaneously at least one measurement for which the results are largely random.}''\\

and \\

``\emph{Decoherence: what happens to quantum superpositions when an attempt is made to distinguish previously indistinguishable ways an event can be realized. It renders superpositions of probability amplitudes into superpositions of classical probabilities. Decoherence has no analogue in classical physics.}''\\

\noindent are therefore questionable as there is no proof that randomness is an irreducible characteristics of quantum physics. In the first assertion, the quantum uncertainty associated to the position and momentum of a particle should not be considered as evidence of the probabilistic nature of quantum mechanics. Position and momentum are concepts that belong to classical physics but loose relevance and rigor at the quantum scale, especially on the level of single particles. In the second assertion, the reason for randomness is in the simplifying assumption that consists in adding a thermal or quantum noise in the modelling (e.g., some leakage in a source or in a beam, or during the detection). Formally, it involves a projection technique analogue to that used when applying Born's rule. These assertions support non-locality as an essential feature of quantum mechanics and hence lead to incongruities such as, e.g., the EPR paradox \cite{Einstein1935}, the quantum eraser paradox \cite{Bracken2021} or Wheeler's delayed-choice experiment \cite{Wheeler1978}, since all the literature associated to the second revolution assumes the reality of a single detection event without accounting for indistinguishability and long time detection. 

With the Born rule remaining at the heart of most theoretical approaches in quantum physics, the dichotomy between the object of study and the measurement device persists; and it is legitimate to ask why a detector during measurement is not described as a quantum system together with the object of study. Moreover, for single microscopic event (or individual) measurements, the use of this rule is questionable. 

\subsection{Redefinition of the measurement without the Born rule} 
Any measurement is in essence a macroscopic process and the use of the Born rule is unnecessary as long as a suitable description of the detection process is done. Generally, two key aspects of measurement are neglected when performing an experiment and analyzing the data. 

First, the identification and spatial isolation of an elementary particle or an atom is impossible to achieve since being either a boson or a fermion, it is indistinguishable from other identical particles or atoms in an experiment, especially when its energy is below that which is required for an excitation process in a detector to be effective. Typically, the threshold is of the order of 1 eV to excite an electron in a semiconductor-based detector. Note that the so-called \emph{indistinguishability} of identical particles (including composite ones like atoms or ions) is one way to consider that ultimately only quanta or excitations of quantum fields exist with specific quantum numbers in contrast to  point-like particles that could be tagged individually. 

To illustrate this important point, we consider two localized states and rewrite them as a delocalized state. For example, suppose two local points $1$ and $2$ where we define the cat state: 
$|\Psi \rangle= [(\hat a^\dagger_1)^2 -(\hat a^\dagger_2)^2]|0\rangle/2$ 
where $\hat a^\dagger_i$ is a creation operator at point $i=1,2$ acting on the vacuum state $|0\rangle$. We immediately see that the rewriting:
$|\Psi \rangle= ({\hat a^\dagger_1} +{\hat a^\dagger_2})({\hat a^\dagger_1} -{\hat a^\dagger_2})|0\rangle/2$ can be interpreted as a factorized combination of two orthogonal  symmetric and anti-symmetric delocalized single photon states. Conversely, the state $|\Psi \rangle= {\hat a^\dagger_1}\hat a^\dagger_2|0\rangle =  
[({\hat a^\dagger_1}+ {\hat a^\dagger_2})^2-({\hat a^\dagger_1}-{\hat a^\dagger_2})^2]|0\rangle/4$ can be seen as two localized separated states or a cat state of two delocalized states. This observation is a consequence of indiscernability and enforces the use of statistics to describe a quantum state as a superposition of different elementary particle numbers. Even in the absence of an interaction force, bosons (fermions) in beams attract (repel) each other resulting in (anti)bunching effect as a result of their mutual quantum  correlations. Therefore, an essential improvement in a theoretical description is to use a multi-particle and multi-mode state in any scattering process rather than the usual amplitude probability associated with a state containing  a determined particle number. Even if in an experiment bosons would be separated far apart from each other in an anti-bunching (or classical) regime, their indistinguishable nature remains and guarantees the impossibility of separating a quantum state into a single particle wave-function in a unique manner as established from the quantum field formalism \cite{Loudon,Huang}. 

Second, in practice one hardly ponders over the significance of measuring an observable whose prediction by quantum mechanics is such that its mean-square deviation provides an error bar comparable to its average (expected value). Normally, one measures quantities with a relatively small confidence interval, otherwise one cannot assess the reliability of a measurement process outcome. Since any EPR or Bell states experiment leads to noisy outcomes, the feasibility of a detection that discriminates between the absence or the presence of a particle raises questions (which will be discussed further below). An avalanche diode works realistically once the uncertainty of having more than one photon interacting is lower than the uncertainty associated with the photon coming from the ambient thermal noise. So, we need the window time $T$ to be large enough to reduce the uncertainty associated with the photon number entering into the detector. For these reasons, we propose that approaches based on the Born rule be replaced by another more pragmatic rule:\\ \emph{A measurement is reliable if its outcome has a quantum uncertainty that is much smaller than the uncertainty associated to its intrinsic noise, mostly the thermal detector noise.}

As a result of the two observations discussed above, there is no need to determine a probability amplitude anymore but instead any macroscopic quantity related to the multi-particle state. Since statistics are necessary but can be performed without use of any probability or stochastic approach, one may suggest alternatively that any experiment is described in terms of macroscopic quantities as done in thermodynamics. A macroscopic quantity such as an accumulated charge due to an electric current, falls in this category. As a consequence, an accurate detector description would include the quantum process of any signal amplification to a macroscopic quantity involving a large ensemble of particles whose relative fluctuations shrink to zero in the thermodynamic limit. A macroscopic approach implies the use of many quantum modes involving a superposition of Fock states. This statement is also valid for any higher-order moment of any macroscopic observable like the fluctuations of particle number or macroscopic current. On the contrary, considering one mode, the particle number and its energy are fluctuating. A particle mode interacting with a detector spreads fluctuations inside the detector rendering a scattering less effective and more difficult to model and to interpret. In the EPR experiment, we show further below that it is the difference between two outcomes with negligible fluctuations that appears to be measurable, thus rendering the quantum measurement local again. 

\subsection{Deterministic view of quantum mechanics}
The renouncement to a probabilistic interpretation as an intrinsic feature of quantum mechanics leads to reassess more firmly the deterministic view of quantum mechanics albeit with a different interpretation from the one given in the EPR experiment \cite{Aspect1981,Aspect1982a,Aspect1982b,Haroche2007,Zeilinger}. But in contrast to classical mechanics, in a deterministic view, any quantum observable like position, momentum or particle number is not well defined for microscopic systems, and so far only the quantum wave function makes sense. Note though that they can be used in a macroscopic or coarse grained sense when the relative uncertainty can be made sufficiently small. In this sense, a precision in the position of $10^{-2}$\AA~(as in an electronic microscope experiment) is possible for a macroscopic object but not for an individual atom, unless it is chemically bonded to the former (though while still individual as an impurity, it is no longer isolated). Probability distributions at the microscopic level should therefore be interpreted only in the statistical sense, i.e. viewed as a histogram over many events.
 
In this deterministic paradigm, the principle of realism amounts to saying that the measured macroscopic quantity has vanishing fluctuations. It sounds {\it a priori} idealistic that a perfect exact measurement requires an infinite number of interrogations but, by analogy, it sounds as idealistic to assume that space and time are continuous since one never approaches an infinitesimal distance or time increment. One needs only to accept that, in any modeling, concepts are defined only in some unreachable limit Leggettcases; otherwise, this would add another layer of unwanted complexity preventing a simple understanding of a physical phenomena. This corroborates the non-cloning theorem proposed by Zurek \cite{WoottersZurek1982} on the impossibility to clone any unknown quantum state justifying the use of many copies to gain knowledge of this state before reproducing it. We therefore propose that: 
{\it Quantum mechanics predicts deterministically that the humankind cannot predict everything}. 

By abandoning the Born rule, the quantum theory becomes complete with no quantum randomness. In fact, incompleteness is rooted in the addition of Born's rule as a postulate, which imposes itself in quantum mechanics as relevant and useful to interpret experimental data using the unjustified simplification that a macroscopic experiment is the result of many separable individual experiments. If this assumption seems to work in practice, it neglects the intrinsic character of indistinguishability of identical particles leading to correlations between the events. 

\subsection{The reality of individual measurement}
Ideally one seeks to optimally design experimental setups to mitigate risks of measurement errors and minimize measurement uncertainty. This problem is typically resolved by carrying measurements over a large number of particles with beams of large number of photons to reduce the fluctuations over the measurement. But foundational experiments \cite{Aspect1981,Aspect1982a,Aspect1982b,Grangier,Haroche2007,Zeilinger,Aspect2015} studying individually one photon or an atom together with separate measurement events each identified as a ``click'' manifestation in a detector, suggest that a second revolution in quantum mechanics has taken place as one can infer from these experiments that quantum systems can be controlled at the individual level. One easily overlooked matter in this picture is that the so-called ``click'' signal corresponds in reality to a picture where an amplification of a single-particle signal has been transformed into a macroscopic outcome, as contrary to the latter the former cannot be detected by a human being. Another overlooked point is that, as shown further below, statistics over many events are always used in these ``individual'' experiments. Claims that each of these events is spatially and timely independent and quantum uncorrelated from each other contradict basic quantum features like uncertainty between position and momentum, as well as over time and energy, and the indistinguishability of identical particles. Such claims need to be rigorously examined.

\section{Continuous measurement versus single-event detection}

\subsection{Detectors as quantum non-Markovian noise removers}

A detector is designed to operate in such a way that the noise of any signal it receives can be reduced or, ideally, suppressed; but the working principle of a detector must also account for the many-body features of an incident beam of photons or atoms. Since a detector is a macroscopic object made of a large number of atoms, the accuracy of any modelling and the validity of the assumptions made should be rigorously assessed. Even though the measurement probability postulate is used in practice in most cases for large numbers of events with the benefit of avoiding  a detector modelling, it raises questions about its validity for individual particle detection. A detector does not measure a probability of occurrence; it measures if an event has occurred or not with an uncertainty related to its intrinsic ``imperfections''. In particular, for a perfect avalanche diode, realism imposes that any detected quantum photon state does not contain the vacuum state in its superposition. This excludes \emph{de facto} a photo-detector which transforms by amplification a quantum superposition of a vacuum state and a one-particle state into a cat state superposition of a vacuum state and a many-particle state. If such a process is possible in quantum mechanics through a suitable unitary transformation, it amplifies uncertainty as well. This apparent paradox is overcome by taking into account the spatial extension of a multi-particle and multi-mode wave packet entering into the detector. The detector becomes sensitive once a sufficient amount of a continuous quantum signal has interacted with it to ensure an uncertainty sufficiently small for an effective recording. Therefore, its working principle is non-Markovian not only in the classical sense but also in the quantum sense. Even if the accuracy can be made as high as possible, an unavoidable uncertainty on the particle number or over its energy is always present since a detector is a localized open system. This results in a quantum superposition over particle number and/or energy states.

The detectors models we propose in the following sections, are quite basic to remain tractable but they clearly show how they differ from single-pulse detection modeling. 

\subsection{Thermodynamics interpretation of the source}\label{therm}

Since any particle beam source is always multi-mode and multi-particle, it is crucial to use a quantum field approach for a rigorous analysis of the detection process. Quite generally, the full quantum characterization of a source is impossible since it should be described within the framework of a Fock space by a global many-particle wave function $|\Psi \rangle $. For example, a source of gamma-ray is not point-like but a sample of matter made of a large number of radionuclides emitting gamma photons from wherever they are within the extent of the sample. If  a spherical wave from the source can be expected on a statistical average, imperfections due to the finite spatial extension of the source generate a complex superposition of photon states that reveals by interference a localized surge of one or more particles that a detector can subsequently detect unambiguously (like in a bubble chamber or on a photographic plate for instance). Note that for high-energy particles, the uncertainty over the position is in principle very small and makes possible to treat the trajectory classically \cite{Mott}.

Any particle is described in terms of its quantum field, which for the photon case, around a reference frequency $\omega_0$, corresponds to the  creation-annihilation operators in the Heisenberg picture
$\hat a^\dagger_\epsilon (x,t)$  and $\hat a_\epsilon (x,t)$  \cite{Navez2023}, restricting ourselves to the $1+1$ dimensional case of position $x$ and time $t$ for each horizontal or vertical polarisation $\epsilon=H,V$. For a free photon field (we can generalize our reasoning for the more complicated case of a massive particle and/or fermion), they obey the commutation relation on the light cone:
$[\hat a_\epsilon (x,t),\hat a^\dagger_\epsilon (x',t')]=\delta(x-ct-(x'-ct'))$. The first-order and second-order correlation functions defined here as:
\begin{eqnarray}
\nonumber
g^{(1)}_\epsilon(x,t,x' , t') &=& \langle \Psi |\hat a^\dagger_\epsilon (x,t) \hat a_\epsilon (x',t')|\Psi \rangle\\
\nonumber
g^{(2)}_\epsilon (x,t,x' , t') &=& \langle \Psi |\hat a^\dagger_\epsilon (x,t)\hat a^\dagger_\epsilon (x',t') \hat a_\epsilon (x,t) \hat a_\epsilon (x',t')
|\Psi \rangle
\end{eqnarray}
are the source characteristics that are usually known. These correlations can be determined experimentally from statistics over a large particle number. Ideally, a full description should involve higher-order correlations or the photon number per each mode. But even this information would be of no avail to address individually each  particle as a consequence of the non-cloning theorem. 

In the following, we use the simplification $\hat a_\epsilon (x,t)=\hat a_\epsilon (s)$ where $s=x-ct$ is the proper distance, and to proceed, we restrict the spacetime over the pulse interval $L=cT$ with $T$ being the time interval of the localized wave, and $s\in[0,L]$. We then define the global operator $\hat a$:
\begin{eqnarray}\label{global}
\hat a= \sum_{\epsilon=H,V}\int_0^L ds \, 
\psi_\epsilon(s)\hat a_\epsilon (s)
\end{eqnarray}
where we introduce the distribution function $\psi_\epsilon(x- ct)=\exp[i\omega_0(x/c -t)]f_\epsilon(x- ct)$, the quantity $f_\epsilon(s)$ being the localized envelope, which satisfies the normalization condition:
\begin{equation}\label{1}
 \sum_{\epsilon=H,V} \int_0^L ds |f_\epsilon(s)|^2=1  
\end{equation}
We also define the Fourier transforms:
\begin{eqnarray}
\hat a_{k,\epsilon}&=& \frac{1}{\sqrt{L}}\int_0^L ds e^{-i(\omega_0/c + 2\pi k/L)s}\hat a_\epsilon (s) \\
\hat a_\epsilon (s)&=& \frac{1}{\sqrt{L}}\sum_{k=-\infty}^\infty  e^{i(\omega_0/c + 2\pi k/L)s}\hat a_{k,\epsilon}\\
f_{k,\epsilon}&=&\int_0^L ds \, e^{-i2\pi k s/L} f_\epsilon(s) \\ 
f_\epsilon(s)&=&\frac{1}{L}\sum_{k=-\infty}^\infty  e^{i2\pi k s/L} f_{k,\epsilon}
\end{eqnarray}

We now focus on the modelling of the radiation considering the following three typical states with very small relative particle number fluctuations:
\begin{enumerate}
    \item 
a Fock state with $N$ particles $|N \rangle = (\hat a^\dagger)^N |0\rangle /\sqrt{N!}$; 

\item
a coherent state $|\alpha \rangle = \exp(\alpha \hat a^\dagger -\alpha^*  \hat a)|0\rangle$ with $N=|\alpha|^2$; 

\item a thermal state with a density matrix 
\begin{eqnarray}\label{th}
\hat \rho = \frac{1}{Z}\exp\left[-\beta\sum_{\epsilon=H,V}\sum_k E_k \hat a^\dagger_{k,\epsilon} \hat a_{k,\epsilon} \right]
\end{eqnarray}
where we define $\beta$ as the inverse of an effective temperature, and where $Z$ is the normalisation coefficient. For this last case,
we can identify the Bose-Einstein distribution and relate the effective energy $E_{k,\epsilon}$ to the average total particle number $N= \sum_\epsilon \langle \hat N_\epsilon  \rangle$. Using the Fourier transform, we find in the weak occupation limit:
\begin{eqnarray}
\langle \hat a^\dagger_{k,\epsilon} \hat a_{k,\epsilon} \rangle\!\! &=& \!\!N|f_{k,\epsilon}|^2\!\!=\!\frac{1}{\exp(\beta E_{k,\epsilon}) -1}\!\simeq\! \exp(-\beta E_{k,\epsilon})\\
\hat N_\epsilon\!\! &=&\!\! \sum_{k=-\infty}^\infty\, \hat a^\dagger_{k,\epsilon}  \hat a_{k,\epsilon}\\
\langle \hat N_\epsilon  \rangle\!\!\! &=& \!\!\!\!\! \sum_{k=-\infty}^\infty\!\! \frac{1}{\exp(\beta E_{k,\epsilon}) -1}
\! \simeq \!\!\! \sum_{k=-\infty}^\infty\!\! \exp(-\beta E_{k,\epsilon}) 
\end{eqnarray}
Quite generally, even using a thermal model where fluctuations are larger than the mean particle number, a deterministic detection is possible with some certainty provided the time window for detection is large enough. The uncertainty is determined from the fluctuations:
\begin{eqnarray}
\langle \delta^2 \hat N_\epsilon \rangle =\sum_{k=-\infty}^\infty \frac{\exp(\beta E_{k,\epsilon}) }{[\exp(\beta E_{k,\epsilon}) -1]^2} \simeq \langle \hat N_\epsilon \rangle
\end{eqnarray}
This allows to conclude here that the relative fluctuations scale like 
$\sqrt{\langle \delta^2 \hat N_\epsilon \rangle}/\langle \hat N_\epsilon \rangle \sim 1/\sqrt{N} \sim 1/\sqrt{L}$ and therefore always vanish in the thermodynamic limit. This is the consequence of an average over many modes (like in thermodynamics). The exception is the case of a mono-mode light corresponding to the Bose-Einstein condensation of light with large fluctuations: 
$\sqrt{\langle \delta^2 \hat N_\epsilon \rangle}/\langle \hat N_\epsilon \rangle \sim 1/N  \sim 1/L$, which is not considered here \cite{Navez1997}. 
\end{enumerate}

\subsection{Scattering approach}
Quite generally, a photon detector can be modelled using scattering theory by analyzing the transition rate of excited electrons using the Fermi golden rule. This rule, which originates from the scattering theory using the S--matrix formalism, imposes an energy conservation between the initial and final states. Although these states look like non-localized states in space such as a pure monochromatic wave, the $i0^+$ prescription appearing in the Lippman-Schwinger equations guarantees an adiabatic switching on of a scattering process and imposes that any signal which is interacting with a detector should be causal, localized in time and in space, without any specific characterization of the extension of the wave function. In fact, a localized wave function always has an uncertainty in energy $\Delta E$, which implies to write it as a superposition of delocalized monochromatic input states, which are subsequently transformed into a localized superposition of delocalized output states under the action of S-matrix. As a result of this localisation, the uncertainty principle fixes a lower bound for the time of scattering given by the pulse time $t_d \sim \hbar/\Delta E$, which differs from the collision time related to the range of interaction. 

Note that we should be cautious to not use the phraseology of transition probabilities as defined for the formal scattering approach \cite{Ryder}, as it is valid only once statistics over many events are considered. In contrast for a unique event, the mere assumption of a superposition over many eigenenergy states reestablishes the reality of a time evolution of a localized wave packet for which these rigorous scattering concepts become useless. With this approach, there is a misconception that in a detection of a set of many successive $N$ fainted photon pulses (as for, e.g., a coherent state of very small amplitude, i.e., $|\alpha|^2 \ll 1$), well separated spatially, they can be assumed independent. Usually, Born's rule is used to assign a probability of detection for each separated pulse. But since an interaction time can be very long in a scattering process, a photon detection is the result of the accumulation of these many coherent pulses in these detectors as the total intensity $N|\alpha|^2$ contains a sufficiently large number of photons. Therefore, the detection process is non-Markovian on the quantum level i.e., a detector cannot reset a virtual state between two successive pulses (unless it is done naturally after a fainted pulse has undergone many reflections) and the single photon detection is a non-instantaneous process resulting from an accumulation of many fainted pulses and not from a probability of occurrence. 

\subsection{Determinism in photo-detector models}
Quite generally, the dependence of the detection efficiency on the (photon or atom) beam intensity is hard to establish especially at the typical low counting rate of 100 photons per second. This raises questions as to the right approach for the calibration of a photo-detector for a single-photon detection. Generally, the black body radiation serves as an intensity reference for the calibration of a photo-detector, which is assumed to have the same sensitivity for any quantum signals such as the  thermal radiation or the fainted coherent pulse containing much less than one photon on average. But, the assessment of this assumption requires a microscopic modelling of the detector establishing the transformation process of a single particle signal into a macroscopic current or photo current \cite{Navez2023}.

In the scattering interpretation, there are two cases: 

\begin{enumerate}

\item {\bf The continuous measurement:} We assume a large length $L$ signal with a large average $\langle \hat N \rangle \gg 1 $ where a detector is in a regime of uncertainty lower than the uncertainty associated to the signal's noise. The modelling is easy since it simply requires to transform a radiation into a continuous electrical current that is subsequently amplified. The scattering formalism is well suited for a continuous process even for an extremely weak signal as  a result of no upper limit for the interaction time \cite{Navez2023}.

\item {\bf Geiger counting:} In high energy regimes, here $\geq$ 1 keV, the uncertainty over the energy and hence over the momentum is generally large enough for a particle to be described classically along its trajectory, which can be followed. For instance, an energetic electron or a photon can, along its path in a material medium, excite several thousands of electrons, which can be subsequently analyzed statistically \cite{Ouerdane2010}. However, at low energy, here $\sim 1$ eV, it must be ensured that the detector ``click'' corresponds to one photon only or a defined distribution for the wave function over a window time interval $T$. The discrete measurement is carried out by avalanche photo-diodes. These detectors are difficult to model at the quantum level due to their intrinsic instability. At least, one can assume that the detection is triggered once the certainty for the quantum state of not being the vacuum state, is ensured during the time $T$.

To give some ideas, we propose three models with different assumptions, showing their non-markovian characteristics:

{\bf Model 1:} Considering a coherent input radiation $|\alpha \rangle$ with an average photon number $\langle  \hat N_\epsilon \rangle \sim 1 $ and polarisation $\epsilon$, the quantum superposition allows the possibility of including the zero-photon state. Adopting a pragmatic view, we impose that the quantum uncertainty over the absence of photons, measured by $|\langle 0|\alpha \rangle|^2=\exp(-\langle \hat N_\epsilon \rangle)$, is lower than the detector noise or equivalently the inefficiency $1-\eta$ where $\eta$ is the quantum efficiency. As a result, we obtain the working  condition $\exp(-\langle \hat N_\epsilon \rangle)< 1-\eta$. For $\langle \hat N_\epsilon \rangle \sim 2$, the efficiency can reach $85\%$. Such model of a detector is really rudimentary as it neglects its quantum internal dynamics and the possibility of photon losses.  

{\bf Model 2:} Consider a radiation that induces a Rabi $\pi$-pulse in the Bloch sphere necessary for the detector to operate and excite exactly one electron for an avalanche in the case of a diode, and for a photo-sensitive reaction in the case of a photographic plate. The Rabi pulse time may be smaller than a $\pi$-pulse leading the detector in a state superposition of ground and excited electronic states after interaction with the pulse. The detector works like a ratchet, meaning that many weak pulses produce many small rotations in the Bloch sphere before the avalanche takes place once a complete rotation has been achieved from the south to the north pole. For an effective rotation angle $\theta$, $\pi/\theta $ pulses are necessary to achieve an avalanche. Such a quantumly non-markovian  detector memorizes the rotation angle after each pulse with some decoherence occurring between each pulse, thereby reducing effectively this rotation angle. As a consequence, if the time interval between two consecutive pulses is too long, the pulse may scatter away before the next one occurs preventing further rotation and hence the counting rate. An example of such detection process is illustrated in Fig.~\ref{fig1}.

{\bf Model 3:} The model 2 has the disadvantage that the final state has a well-defined energy in contrast to a semiconductor avalanche detector involving the transition of many electrons between the valence and the conduction bands in the continuum of the wavevectors space. The Fermi golden rule tells that a set of successive distant fainted pulses is  decomposed into its wavenumber components as the respective channels of interaction for an effective excitation to take place. The process is therefore collective as it involves many non-separable pulses. In presence of radiation, the conduction electrons number $N_{c}$ obeys the Einstein equation:
\begin{eqnarray}\label{Einstein}
\frac{d N_{c}}{dt}= - A N_c +  B u (N_v- N_c) 
\end{eqnarray}
where $A$ and $B$ are the Einstein coefficients for the spontaneous and stimulated emission respectively usually calculated using Fermi's golden rule and $u$ is the  spectral energy density per unit bandwidth given by $u=\sum_\epsilon \hbar \omega_0 \langle \hat N_\epsilon \rangle t_d/ (L l^2)$ with $t_d$ the detection time (related to $L=c t_d$ for photons) and $l$ the detector size. In thermodynamic equilibrium, the Bose-Einstein statistics associated to the black body radiation must be recovered, which in turns imposes that  $A/B=  \hbar \omega^3_0/\pi^2 c^3$. Using similar thermodynamic arguments, the fluctuations $\delta^2 N_{c}$ obey a dynamics similar to that of $N_c$: 
\begin{eqnarray}
\delta^2 N_{c}\sim  N_c 
\end{eqnarray}
For a weak constant photon flux and $N_c \ll N_v$, the spontaneous and stimulated emission rate happen to be negligible. The solution of Eq.~\eqref{Einstein} for  the relative fluctuations scales like $\sqrt{\delta^2 N_{c}}/N_{c} \sim 1/\sqrt{B t\sum_\epsilon \langle \hat N_\epsilon \rangle}$, which goes to zero very quickly for large time $t$ guaranteeing a deterministic detection. We can again fix an avalanche  detection threshold for $N_{c}$ on the order of unity which can occur after a very long time $t_d$ related to the size of the signal. 

Another limitation is the inter-distance between the energy levels $\sim \hbar^2 /m l^2$  within the bands which provides an upper bound to the pulse duration $t_d \sim l^2 m/\hbar$, which is also the detection time. It renders Fermi's golden rule for estimating $B$ inadequate since the energy band continuum assumption is not valid and would not make the energy transition effective. This point can be illustrated with a simple example. For a photo-diode of cubic size $l=1$ mm and $m=10^{-30}$ kg, we find the upper bound $t_d \ll 10$  ms. For larger semi-conductor detectors, this bound becomes even larger. Beyond this bound, the model 2 applies with a decay rate that becomes significant, hence precluding an efficient single-photon detection. Given this time scale and a radiation wavelength of about $\lambda_0=2\pi c/\omega_0 \sim 1\mu m$, we infer from the condition of low spontaneous emission $N_v \gg (N_c /u) (A/B)$ with $N_c=1$ that the minimal carrier density $n_v=N_v/l^3$ in the valence band necessary for an effective transition is:
\begin{eqnarray}
n_v \gg \frac{\omega^2_0}{\pi^2\langle \sum_\epsilon \hat N_\epsilon \rangle l c^2}=\frac{4}{\langle \sum_\epsilon \hat N_\epsilon \rangle l \lambda_0^2}\sim 4\times10^{9}\mbox{cm$^{-3}$} 
\end{eqnarray}
Typically the carrier density is in the 10$^{18}$ cm$^{-3}$ range in semiconductors. The existence of such bounds on time and carrier density hinders the isolation of single events well separated in time and with too weak intensity.
\end{enumerate}

These models show the major limitations of the detectors capabilities, which make it practically impossible with the current available technology to isolate at low energy one pulse event from another in order to control single photons well-separated in time. Moreover, one common bias in the analysis of experiments is to expect that if the statistical distribution of particles in a beam is Maxwellian, the particles behave independently and classically while even in this case, indistinguishability still plays a role during detection. However, indistinguishability cannot be directly observed as it involves the detector itself, and one would need a second detector measuring the state of the first one before it measures any signal (but to our knowledge, such a realization has not yet been implemented). Hence, the individual event approach in quantum mechanics works only in the statistical sense. As a result, the advantage of using an avalanche detector in comparison to a continuous detector easier to model, becomes less obvious. The above modeling approach changes totally the perception of the apparent paradoxes in quantum mechanics, in particular, Schr\"odinger's cat and the Bell experiment, discussed below.

\section{The Schr\"odinger's cat paradigm}
\subsection{Coherent state experiment}
The most common state is a coherent state whose uncertainty in the photon number is already present. This state prevents a genuine superposition of a polarisation state as any prism would factorize the resulting state into two disentangled ones:  
\begin{eqnarray}\label{Schr}
|\alpha \rangle &=& |\alpha_H \rangle_H |\alpha_V \rangle_V \\
\nonumber
|\alpha_\epsilon \rangle_\epsilon & = &\exp \left(-|\alpha|^2\int_0^L ds \, |\psi_\epsilon (s)|^2/2 \right)\\
&\times& \exp\left(\alpha  \int_0^L ds \,  \psi_\epsilon (s)\hat a^\dagger_\epsilon (s) \right) |0 \rangle_\epsilon
\end{eqnarray}
\noindent The common confusion leading to  the Schr\"odinger's cat paradox stems from the use of Born's rule to project Eq.~\eqref{Schr} onto a one-particle entangled state, which, assuming a 50\%/50\% beam splitter, reads
\begin{eqnarray}\label{E}
|\psi\rangle_E & = & (|0\rangle_H |1 \rangle_V + |1\rangle_H |0 \rangle_V)/\sqrt{2}\\
|1 \rangle_\epsilon & = & \frac{1}{{\cal N}_n} \left[\int_0^L ds \,  \psi_\epsilon (s)\hat a^\dagger_\epsilon (s)  \right]|0 \rangle_\epsilon
\end{eqnarray}
where ${\cal N}_n$ enforces the normalisation. The state $|\psi\rangle_E$ is the result of a low particle number approximation $|\alpha| \ll 1$, which as a consequence creates a fainted pulse mostly consisting of the vacuum and a one-particle states (states with larger particle numbers being negligible). Equation~\eqref{E} is an entangled state contrasting with the former, Eq.~\eqref{Schr}, which is disentangled in the polarization.  

This fundamental difference shows the pitfalls of this projection approximation leading to conclusions on the intrinsic randomness in quantum mechanics and, as a result, to the cat paradox with the measurement outcome deciding about the life or death state.  Had such an approximation not been done, the conclusion would be far different. Instead, treating $\psi_\epsilon (s)$ as a random variable but with a constant total intensity satisfying Eq.~\eqref{1} and imposing the criterion that $\int_0^L ds\, |\psi_\epsilon (s)|^2 >1/2$ for a photon to be detected in an ``avalanche-like'' detection in the $\epsilon$ polarisation channel, the paradox is then explained in terms of lack of a perfect  control of the photon beam. Thermal fluctuations may preclude the preparation of a beam with a perfect polarization. As a consequence the parameters characterizing the functions $\psi_\epsilon (s)$ are the ``hidden variables'' that could tell in advance the outcome for the cat: being alive or dead. Similarly to what we discussed concerning the detector models in the previous section, a set of many pulses containing at least one photon is sufficient for reaching the detection threshold even if each pulse has individually less than one photon.  

\subsection{The genuine single photon generation}
\begin{figure}
\begin{center}
 \includegraphics[width=0.4\textwidth]{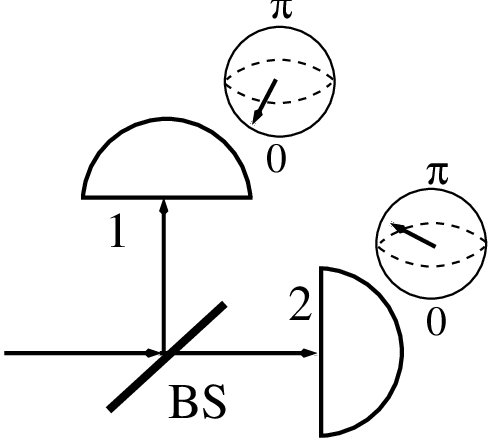}
\end{center}
\begin{tabular}{|c|c|c|c|}
\hline
\multicolumn{2}{|c|}{Detector 1} & \multicolumn{2}{c|}{Detector 2} \\
\hline
state & outcome & state & outcome \\
\hline
$0$ & $0$ &$\pi/2$ & $0$ \\
$\pi/2$ & 0 & $\pi$ or $0$ & $0$ or $1$ \\ 
$\pi$ or $0$ & $0$ or $1$ & $3\pi/2$ or $\pi/2$ & $0$ \\
$3\pi/2$ or $\pi/2$ & $0$  & $0$ or $\pi$ & $1$ or $0$ \\
\hline
\end{tabular}
\caption{Example of non-Markovian detection based on model 2: Sequence of four $\pi/2$ pulses entering two avalanche detectors, resulting from four separate single photon signals passing through a 50/50 beam splitter (BS). Determinism imposes the absence of a quantum reset, which leads to a memory effect. Once the detector is in a $\pi$ state after the electronic transition, the avalanche is effective or not, according to the detector random noise state. If the two detectors are synchronized, the outcomes are correlated leading to coincidence measurement outcomes. If one detector state is shifted by $\pi/2$ relatively to the other which is more common on average, the outcomes are anti-correlated with no coincidence (see table) and corresponds to the perfect prediction of one photon per pulse if the avalanche is always effective. More generally, the impossibility of the detector to tell its own state due to its noise and the presence of many electrons (see model 3), instead, prevent any synchronisation causing the randomness of both single and coincidence outcomes such as in Ref.~\cite{Grangier}. In contrast, a model detector based on Born's rule would have produced 0/1 or 1/0 outcomes with probability 1/2. In this example, the decoherence time  for the detector  to reset or not to its ground state  between two events (like two separated pulses of light) is one factor responsible for this randomness.
}
\label{fig1}
\end{figure}

Using a fainted coherent pulse seems a practical way of producing an isolated photon but justifying the absence of effects of higher photon numbers in the superposition of states, is an issue. By exciting a single ``artificial'' atom by means of a Rabi $\pi$-pulse from a classical radiation source, a subsequent spontaneous emission will produce a real single-photon pulse with a broadening given by the decay rate. Another way to produce one is to use entangled beams where the measurement of one photon in one beam ``guarantees'' the presence of one photon in another beam \cite{Grangier}. Owing to the precision of an avalanche diode; however, one may question the suitability of such a post-selection method once it is established that there cannot be separability between the events. If this process is performed within a cavity, the single photon has a channel for producing a genuine superposition leading to a pure entangled state using a beam splitter: $|\psi\rangle_1 =(c_V|0\rangle_H |1 \rangle_V + c_H|1 \rangle_H |0 \rangle_V)$ with $|c_V|^2+ |c_H|^2=1$ as shown in Fig.~\ref{fig1}. Quantum cloning is in principle possible in a known basis and therefore in the Fock basis. If we suppose that an ideal detector leads to the cat state from amplification: $|\psi\rangle_n =(c_V|0\rangle_H |n \rangle_V + c_H|n \rangle_H |0 \rangle_V)$, then the relative fluctuations are very large:
$\sqrt{\langle \delta^2 N_\epsilon \rangle}/{\langle  \hat N_\epsilon \rangle } =1/\sqrt{|c_\epsilon|^2 (1-|c_\epsilon|^2)}$, and do not shrink for large cat sizes. But for a realistic case, such states do not appear to be created easily experimentally since all fluorescence photons cannot be detected individually; once we admit that the detection is not a markovian process.  

\subsection{Quantum non-demolition measurement}

\subsubsection{Superconducting qubit inside a cavity}
Quantum non-demolition (QND) measurement is one of the current techniques \cite{Braginsky1996} for monitoring few-level systems (qubit or cavity photon). In a quantum circuit, a qubit readout is achieved by an interaction with a detuned cavity allowing a quantum non-demolition measurement in the dispersive regime. Concretely, the qubit readout involves the measurement of a macroscopic current through a microwave waveguide and uses the Stark shift effect to discriminate between one state or another in a specific basis \cite{Martini}. But in the case of a superposition, there is no such detailed experimental analysis providing evidence that the Born rule applies. It may well be that the outcome is such that no discrimination is possible between one eigenstate over another (possibly leading to dark counting) and still applying Born's rule \cite{Devoret2013} while its validity is questionable.

If a qubit is in an initial state $|\psi\rangle_1 =c_0|0\rangle  + c_1|1 \rangle_V$, an external input radiation allows, after interaction, the transmission of an ideal coherent state $|\alpha \rangle$ inside a cavity mode for the excited state,  the final ``cat-like'' state $|\psi\rangle_1 =c_0| 0\rangle |0 \rangle + c_1|1 \rangle |\alpha \rangle$ \cite{Devoret2013,Blais2004a,Wallraff2024}.
The resulting output signal has two possible unambiguous intensity outcomes  $0$ or $I=|\alpha|^2$ in absence of a superposition. Considering $N$ measurements without a detector reset in between two events, the average total intensity is $\overline{I}=N |c_1|^2 I$ and the fluctuations $\overline{\delta^2 I}=N [|c_1|^2(1-|c_1|^2)]I^2$. Clearly, the relative fluctuations shrink to zero for large $N$. In this sense, $|c_0|^2$ and $|c_1|^2$ are the probability amplitude that is measurable after a counting over many events. A more thorough theoretical description of a qubit and resonator should have been used involving a quantum field approach using a output Fock basis as in Section \ref{therm} \cite{Navez2023}.

\subsubsection{Rydberg atom passing through a cavity}
In Ref.~\cite{Haroche2007}, the photon distribution in the state $\sum_{n=1}^\infty c_n|n \rangle$ is probed in a cavity with a beam of atoms in a Rydberg state, each of which being subsequently detected in one of the two possible spin states, $|+\rangle$ and $|-\rangle$, ground and excited, respectively. For each atom individually,  a QND operation is carried out on an initially factorized state  $|\psi \rangle =(|+\rangle + |-\rangle)\sum_{n=1}^\infty c_n|n \rangle$ transformed into $\sum_{n=1}^\infty [(1+e^{i\theta n)})|+\rangle + (1- e^{i\theta n})|-\rangle ]c_n|n \rangle$ where $\theta$ is the induced spin rotation angle after the interaction. Starting from a state superposition, successive interrogations induced by successive measurements favor one particular photon number state. Although this experiment suggests that the outcome of the quantum signal originates from  the random results of measurements and would generate an effective wavepacket collapse \cite{Haroche2007}, the theory behind this interpretation is not based on a ``many-mode'' quantum field theory. Again, the assumption over the independence of atoms interacting with the cavity, as they are  well spatially separated from each other, cannot work when they are in a superposition of many-body states and may well change the prediction on the cat states.

Since atoms are indistinguishable, we need to work in the many-body representation of a quantum atom field in the two possible Rydberg states. The use of the measurement postulate in a two-state basis for a single atom contrasts with a macroscopic population measurement in two states with the aim of lowering the uncertainty by increasing the atom number. For instance, two atoms evolving into an equal superposition of two Rydberg states can be seen from the detector as one atom in state $g$ and another in state $e$ as they are not separable in time. Therefore, the concept of wavepacket collapse is based on the validity of the Born rules to establish the intrinsic randomness of quantum mechanics. A more elaborated model that includes the indistinguishability between atoms would restore the determinism instead and may alter the conclusion established in \cite{Haroche2007}. 

\section{Infeasibility of event separation in the Bell experiments}
\subsection{General considerations}
In its orthodox formulation, quantum mechanics does not support the principles of locality and realism, and a wealth of experimental works inspired by Bell's inequality test and conducted for over five decades \cite{Freedman1972,Aspect1981,Aspect1982a,Aspect1982b,Weihs1998,Rowe2001,Matsukevich2008,Ansmann2009,Scheidl2010,Giustina2013,Christensen2013,Handsteiner2017}, led their authors to rule out both principles. Interestingly, the majority of these experiments involve the polarization direction of light as the physical quantity of interest, while Bell's theoretical work considered 1/2 spins and a Stern-Gerlach-like experiment \cite{Bell1964}. Further, and as important, the cited experiments have loopholes (such as, e.g., detection, locality, memory effects and statistics \cite{Brunner2014}) which preclude absolutely certain conclusions. In 2015, a work reported a loophole-free Bell inequality violation with an experiment involving electron spins \cite{Hensen2015}. The authors conclude that ``[their] \emph{result places the strongest restrictions on local realistic theories of nature to date''}; and this is indeed the case. But one may still argue that the principles of locality and realism are not disproved yet, and that other loopholes waiting to be identified may have been overlooked.

Here, for clarity, we neither claim that quantum correlations or entanglement at long distance are not possible, nor that Bell states do not exist. The singlet or triplet states in the helium atoms are examples of such entangled electron spin pairs. However, performing any Bell-like experiment in a framework based on a Fock space description imposed by quantum field theory is questionable once it is established that the collected data correspond to events that are not well separated because of their non-vanishing space and time correlations. Hence, there cannot be independent measurements in the context of a Bell experiment. This is a physics-based loophole, which never was examined. Such a type of experiment would certainly be feasible if there would be a means to distinguish identical particles, which would substantiate the assumption that two consecutive measurement events can be well separated. But given the experimental reality of non-discrimination between the detector and the setup, the current interpretation of Bell experiments does not involve the physical description of the detector and the source. 

Without the Born's postulate, the quantum description of the Bell experiment needs be reformulated. Particularly, the impossibility of a perfect interpretation of a noisy measurement prevents its feasibility, and hence any of two non-local correlated observed outcomes (spin up, spin down) is necessarily the result of many non-separable events. Under these conditions, locality, causality, and completeness are restored and there is no need of additional, hidden variables as shown below. 

\subsection{Influence of the source thermal noise on photon pair production}
A set of pair states produced in an experiment is quite different from what John Bell assumed in his work \cite{Bell1987}. Using a continuous variable formalism associated to photons, we can factorize the polarized state symmetrically if the  EPR source is not noisy. But in this case, all outcomes in both polarization directions would occur simultaneously contrary to observations where only one direction is observed. A possible explanation for such a discrimination is the inherent thermal noise of the source, like a beam of calcium atoms \cite{Aspect1981,Aspect1982a,Aspect1982b} or a nonlinear crystal \cite{Zeilinger} that generates additional classical photon flux fluctuations blurring the intrinsic quantum ones, creating a set of random pulse events. In a particle picture, an itinerant photon shows a Brownian-like dynamics as it interacts with atoms whose motion relative to the detector produces a Doppler shift with a relative frequency change $\sim \sqrt{1/\beta m}/c \sim 10^{-6}$ where $m$ is the mass of the atoms of the source, $c$ the speed of light, and $\beta$ the inverse thermal energy reflecting the temperature statistically determined from the average over the speeds of the atoms in motion.

In addition to the operator $\hat a_{k,\epsilon}$, we introduce the pair operator $\hat b_{-k,\epsilon}$ both with their specific modes $k,\epsilon$ in one dimension. In the interaction representation, the Hamiltonian describing the source of size $L_s$ has the form \cite{Kolobov1989,Navez2001}:
\begin{eqnarray}\label{H}
\nonumber
\hat H_I(t) &=& \sum_{k, \epsilon, \epsilon'}D_{k,\epsilon,\epsilon'}(t)  \left(\hat a^\dagger_{k,\epsilon}\hat a_{k,\epsilon'}\!+\!\hat b^\dagger_{k,\epsilon}\hat b_{k,\epsilon'}\right)\\
&+& K\sum_{k,\epsilon} \left(\hat a_{k,\epsilon}\hat b_{-k,\epsilon} + \hat a^\dagger_{k,\epsilon}\hat b^\dagger_{-k,\epsilon} \right)
\end{eqnarray}

\noindent where  $D_{k,\epsilon,\epsilon'}(t)=\Delta_{k}+ \eta_{k,\epsilon,\epsilon'}(t)$ is decomposed into
the detuning term $\Delta_{k} =\omega_{k}-\omega_p/2$, which we assume constant in a selected region around the wave number $k_0$, and the noise term $\eta_{k,\epsilon,\epsilon'}(t)$ due the random Doppler frequency variation. The coupling term $K$ generating the photon pairs, results from the input of two lasers exciting the calcium atoms, or from the non linear crystal with one pump laser of frequency $\omega_p$ and amplitude of the pump field $\mathbf{E}$ with an expression scaling like $K \propto |\mathbf{E}|$. It may be time-dependent due to the internal fluctuations of the laser intensity. Under ideal circumstances, the time dependence can be neglected and $\eta_{k,\epsilon,\epsilon'}(t)=0$ for a symmetric system under polarization, which means that pairs are produced in all polarisation with equal weight. The crucial point here is that the single value decomposition $D_{k_0,\epsilon,\epsilon'}(t)=\sum_{j=1,2} A^*_{\epsilon,j}(t) d_{j}(t) A_{\epsilon',j}(t)$ (with $\sum_\epsilon |A_{\epsilon,j}|^2=1$) provides a history and assigns a preferential polarization when the eigenvalues are not equal $d_{1}(t)\not=d_{2}(t)$. 

We can consider it as a stochastic variable whose relative average variation is given by the Doppler shift of a moving atom within the nonlinear crystal or the velocity distribution of calcium atoms. This results in a variance given by $\overline{\eta^2_{k,\epsilon,\epsilon'}}(t)=\delta_{\epsilon,\epsilon'}(\hbar \omega_k)^2 /\beta m c^2$. Comparing with the detuning and the nonlinear coefficient, these stochastic terms have an effect $(\omega_{k_0}/\Delta_{k_0})\sqrt{1/\beta m c^2}$ which can be of the order of unity or greater. As a consequence, such important contributions lead to potentially strong local variations in the intensity for creating histories of polarization mimicking hidden variables.

A basic formulation of the Bell experiment relies on the multi-mode squeezed state 

\begin{equation}\label{Squ}
|\Psi(t) \rangle\!=\! \exp\!\!\left[\!\sum_{k,\epsilon, \epsilon'=H,V}\!\!\!f_{\epsilon,\epsilon'}(t) \hat a^\dagger_{k,\epsilon}\hat b^\dagger_{-k,\epsilon'}\!- \!f^*_{\epsilon,\epsilon'}(t) \hat a_{k,\epsilon}\hat b_{-k,\epsilon'}\!\right]\!\!|0\rangle
\end{equation}
In a steady regime where $D_{k_0,\epsilon,\epsilon'}(t)$ slowly varies in time, we can use the results of Refs.~\cite{Kolobov1989,Navez2001} and, based on the  Hamiltonian (\ref{H}), find  the  connection:
\begin{eqnarray}\label{Squ2}
f_{\epsilon,\epsilon'}(t)=\sum_{j=1,2} A^*_{\epsilon,j}(t) \lambda_{j}(t) A_{\epsilon',j}(t)
\end{eqnarray}
where $\Omega_{j}(t)=\sqrt{d_{j}^2(t)-K^2}$ and with the eigenvalues:
\begin{eqnarray}
\lambda_{j}(t)\!=\!\frac{K\sin(\Omega_{j}(t)L_s/c)} {\left|\cos(\Omega_{j}(t)L_s/c)\Omega_{j}(t)\!-\!i d_{j}(t)\sin(\Omega_{j}(t)L_s/c)\right|}    
\end{eqnarray}

When we trace out over the Fock space associated to the second beam, we recover the form of expression Eq.~\eqref{th} for the thermal density matrix:
\begin{eqnarray}\label{Al}
\hat \rho(t) &=& \frac{1}{Z}\exp\left[\sum_{k,\epsilon,\epsilon'=H,V} G_{\epsilon,\epsilon'} \hat a^\dagger_{k,\epsilon} \hat a_{k,\epsilon} \right]\\
G_{\epsilon,\epsilon'}&=&\sum_{j=1,2} A_{\epsilon,j}(t)\ln|\tanh^2(\lambda_{j}(t))| A^*_{\epsilon',j}(t)
\end{eqnarray}
which guarantees a well-defined average photon number even for a fainted signal, provided many modes $k$ are involved which is the case for a pulse.

Assuming instead that pulses with photon numbers greater than unity are negligible with only one wavevector mode, we could employ the naive approximation of projecting the state  onto the particle number pair of Eq.~\eqref{Squ} to recover the symmetric Bell entangled state:

\begin{eqnarray}\label{Bell}
 |\psi\rangle \sim \frac{1}{\sqrt{2}}\left(\hat a^\dagger_{H} \hat b^\dagger_{H} + \hat a^\dagger_{V} \hat b^\dagger_{V}\right)|0\rangle
\end{eqnarray}
where we use the simplification: $\hat a_\epsilon = \sum_k \hat a_{k,\epsilon}$ and $\hat b_\epsilon = \sum_k \hat b_{k,\epsilon}$. The projection postulate then allows different polarisation outcomes. The expression of the density matrix in Eq.~(\ref{Al}) shows that using a QED formalism, genuine quantum correlations appear among different momentum modes but not among different localizations in space as Eq.~\eqref{Bell} leads to think. Hence the question: if starting from a local quantum state that evolves according to quantum field theory, is it possible to produce a genuine entanglement at two distant points in space resulting in a non-local quantum state? According to \cite{Bricmont}, the existence of such non-local state implies non locality. Most of the textbooks assume their existence but the production of non locality starting from locality is never addressed on a firmly grounded basis.

In comparison using a symmetric state ($\lambda_{1}=\lambda_{2}$) in Eq.~\eqref{Squ}, it is not possible to discriminate deterministically between two states of polarisation which appear now to be factorized and therefore uncorrelated. With our measurement definition where an avalanche diode is triggered once the number of particles exceeds unity, identical outcomes occur for each beam. 

\subsection{The more likely realistic case in experiment}

Once the probability postulate is dismissed, we can establish that the allegedly  ``hidden variables'' originate from the noise of the sources. These thermal asymmetric fluctuations in the source lead to an asymmetry of polarisation in the wave function favoring one polarisation state over another. For a weak non-linearity $K \ll \Delta_{k_0}$, we estimate the relative asymmetry as 
\begin{eqnarray}
\frac{|\lambda_{1}- \lambda_{2}|}
{|\lambda_{1}+\lambda_{2}|}\sim \sqrt{\frac{1}{\beta mc^2}}\frac{\omega_{k_0}}{\Delta_{k_0}}
\end{eqnarray}
For a fainted pulse, if the total intensity $I\sim \sum_{j} \lambda_{j}^2$, then by setting the detector threshold to one half of the total intensity $I_{th}=I/2 $, one obtains a majority rule for knowing what polarisation is measured. According to the noise state of the detector, this threshold may also vary, thus leading to an absence of detection or a double counting as an unavoidable error. Under these conditions, the current experiments on Bell inequality may be well interpreted deterministically with an adequate quantum description of the source and the detector. 
 
\section{Conclusion}
In this work, we critically discussed the Born rule and what its application implies for experimental data interpretation, and consequently for the basic aspects of quantum physics. We argued that once the Born rule as a postulate is dismissed, quantum mechanics can no longer be considered as a probabilistic theory. In fact, this rule is not necessary if a quantum model including both the object of study and the measurement device it interacts with during measurement, based on a scattering approach, is used for data analysis and interpretation. This obviously adds a layer of complexity to the work but with the aim of restoring realism since the existing physical properties of the studied quantum system should not be defined in terms of a probability of occurrence but in terms of macroscopic deterministic quantities. As a consequence, any experiment like the Bell experiment, loses its relevance as a test of the completeness of quantum mechanics, if the assumption on separable single events cannot be fulfilled. Our ignorance of the exact details of the thermal noise in the input state and in the detector state should not imply that quantum mechanics is necessarily probabilistic, but it rather shows the use of probability as a practical convenience. Hence, while Born's rule may be only applicable as a practical tool for experiments involving a large number of particles leading to a statistical and coarse-grained interpretation, it is unsuitable for experiments involving a few qubits or a few cavity photons to be described as individual entities. 

We also discussed with basic illustrative examples the use of the second quantization multi-particle formalism to avoid the pitfalls of the wavepacket collapse in the Schrodinger's cat paradox. While the use of statistics is inherent to quantum physics experiments, this does not necessary entail that probability is a basic feature of quantum mechanics. It is inherent because the intrinsic uncertainty associated to common observables including, e.g., position, momentum, or particle number, requires a large number of data to be properly defined as a quasi-noiseless macroscopic measurement, which is essential for a conclusive interpretation. This amounts to adopting a thermodynamics formulation of the problem, where only macroscopic quantities are measurable since any single-event signal is too noisy to ensure a valuable characterization. For example, the observation of the electronic cloud in a solid crystal under a scanning tunnelling microscope is not a reality proposed by quantum mechanics but a representation of the total electronic density distribution resulting from of a macroscopic measurement current. From our proposed many-body modeling including the thermal noise of the source and of the detector, we conclude that the experiments in Refs.~\cite{Aspect1981,Aspect1982a,Aspect1982b,Aspect2015,Zeilinger} are not proven either to correspond to the idealistic Bell's proposal. Any genuine modeling of a quantum correlation experiment, with the hope of more deterministic predictions, should ideally include:
1) The full formalism of quantum electrodynamics with a many-body formalism highlighting indistinguishability; 
2) A correct modelling of the source (e.g., a calcium beam with the two lasers); 
3) An accurate quantum model of the detector which operates in a finite time and whose finite size imposes a lower bound over the counting rate preventing genuinely distinguishable (time-separated) single-event detection. A procedure along these steps results in a deterministic quantum description including locality and causality.

More complex experiments that necessitate modeling beyond the traditional scattering approach using the S matrix, such as, e.g., the avalanche effect, need also new theoretical methods to address properly the exact quantum dynamics involved. For instance, the formalism of transmission within waveguides usually derived from classical electromagnetism has been extended using quantization \cite{Navez2023}. These developments raise the fundamental questions on the possibility of creating a superposition of macroscopic ``cat'' states leading to large quantum fluctuations and therefore not measurable. In this work, we have shown the difficulty to produce such a ``multiverse'' state given the presently available technology; but a more accurate proof is required to establish this observation on a more formal basis. The non existence of such states amounts to conjecturing that any macroscopic observable has its set of possible eigenvalues restricted to regions within its fluctuations, discarding any extreme value far from its average. Such a statement should ideally be proven starting from a quantum field  model.

As a final remark regarding the discrimination in the theoretical description between the object of study and the measurement apparatus, it is worth recalling that when electric circuits were discovered in the time of Alessandro Volta in 19th century, both the circuit  (e.g. voltaic pile + metal wires) and the measurement device  (e.g. compass or galvanometer) had eventually required the electromagnetic theory for a full understanding of an experiment. 

\begin{acknowledgments}
PN is grateful to Mark Everitt, Gilbert Moutaka for fruitful discussions enabling a deeper contextualisation of the problem. 
\end{acknowledgments}


\noindent 
\begin{thebibliography}{99}

\bibitem{WoottersZurek1982} W. Wootters and W. Zurek, \emph{A single quantum cannot be cloned}, Nature {\bf 299}, 802 (1982).
%https://doi.org/10.1038/299802a0

\bibitem{Dieks1982} D. Dieks, \emph{Communication by EPR devices}, Physics Letters A {\bf 92}, 271 (1982).
%https://doi.org/10.1016/0375-9601(82)90084-6

\bibitem{Born1926} M. Born, \emph{Zur Quantenmechanik der Stoßvorg{\"a}nge}, Zeitschrift f\"ur Physik {\bf 37}, 863 (1926).
%https://doi.org/10.1007/BF01397477

\bibitem{Dirac1930} P. Dirac, \textit{The Principles of Quantum Mechanics} (Oxford University Press, 1930).

\bibitem{vonNeumann1932} J. von Neumann, \textit{Mathematische Grundlagen der Quantenmechanik} [Mathematical Foundations of Quantum Mechanics] (Springer, 1932).

\bibitem{Mermin2017} N. D. Mermin, \emph{Why QBism is not the Copenhagen interpretation and what John Bell might have thought of it}, in: R. Bertlmann, A. Zeilinger, (eds) Quantum [Un]Speakables II. The Frontiers Collection. Springer, Cham. 
%https://doi.org/10.1007/978-3-319-38987-5\_4

\bibitem{Devoret2013} M. Hatridge , S. Shankar, M. Mirrahimi, F. Schackert, K. Geerlings, T. Brecht, K. M. Sliwa, B. Abdo, L. Frunzio, S. M. Girvin, R. J. Schoelkopf, and M. H. Devoret, \emph{Quantum back-action of an individual variable-strength measurement}, Science {\bf 339}, 178 (2013).
%doi: https://doi.org/10.1126/science.1226897

\bibitem{Weschler2021} S. Wechsler, \emph{The quantum mechanics needs the principle of wave function collapse -- but this principle should not be misunderstood},  Journal of Quantum Information Science {\bf 11}, 42 (2021). 
%doi: https://doi.org/10.4236/jqis.2021.111004. 

\bibitem{Luders1950} G. L\"uders, \emph{\"{U}ber die Zustands\"{a}nderung durch den Me{\ss}proze{\ss}}, Annalen der Physik {\bf 443}, 322 (1950).
%doi: https://doi.org/10.1002/andp.19504430510

\bibitem{deBroglie} L. de Broglie, \emph{Ondes et mouvements}, (Gauthier-Villars, Paris, 1926).
%doi:

\bibitem{Bohm1952I} D. Bohm, \emph{A suggested interpretation of the quantum theory in terms of ``hidden'' variables. I}, Physical Review {\bf 85}, 166 (1952)
%doi: https://doi.org/10.1103/PhysRev.85.166

\bibitem{Bohm1952II} D. Bohm, \emph{A suggested interpretation of the quantum theory in terms of ``hidden'' variables. I}, Physical Review {\bf 85}, 180 (1952).
%doi: https://doi.org/10.1103/PhysRev.85.180

\bibitem{Everett1957} H. Everett, III, \emph{``Relative state'' formulation of quantum mechanics}, Reviews of Modern Physics {\bf 29}, 454 (1957).
% doi: https://doi.org/10.1103/RevModPhys.29.454

\bibitem{Cramer1986} J. G. Cramer, \emph{The transactional interpretation of quantum mechanics}, Reviews of Modern Physics {\bf 58}, 647 (1986). 
%doi: https://doi.org/10.1103/RevModPhys.58.647

\bibitem{tHooft2016} G. 't Hooft, \emph{The cellular automaton interpretation of quantum mechanics}, Fundamental Theories of Physics series (Springer Cham, 2016).
%doi: https://doi.org/10.1007/978-3-319-41285-6

\bibitem{Bassi2013} A. Bassi, K. Lochan, S. Satin, T. P. Singh, and H. Ulbricht, \emph{Models of wave-function collapse, underlying theories, and experimental tests}, Reviews of Modern Physics {\bf 85}, 471 (2013).
%doi: https://doi.org/10.1103/RevModPhys.85.471

\bibitem{editorial2022} Survey the foundations. Nature Physics {\bf 18}, 961 (2022).
%doi:https://doi.org/10.1038/s41567-022-01766-x

\bibitem{Heisenberg} W. Heisenberg, \emph{Physics and Philosophy: The Revolution in Modern Science}, pgs. 24–25 (Harper, New York, 1958).

\bibitem{Dirac1927} P. Dirac, \emph{The physical interpretation of the quantum dynamics}, Proceedings of the Royal Society A {\bf 113}, 621 (1927).
%doi:https://doi.org/10.1098/rspa.1927.0012

\bibitem{Jordan1927a} P. Jordan, \emph{{\"U}ber quantenmechanische Darstellung von Quantenspr{\"u}ngen}, Zeitschrift f{\"u}r Physik {\bf 40}, 661 (1927).
%doi:https://doi.org/10.1007/BF01451860

\bibitem{Jordan1927b} P. Jordan, \emph{{\"U}ber eine neue Begr{\"u}ndung der Quantenmechanik}, Zeitschrift f{\"u}r Physik {\bf 40}, 809 (1927).
%doi:https://doi.org/10.1007/BF01390903

\bibitem{proof} As regards some attempts to derive the Born rule, we note that: `` The conclusion seems to be that no generally accepted derivation of the Born rule has been given to date, but this does not imply that such a derivation is impossible in principle '' \cite{Landsman2008}. A more recent discussion is in Ref.~\cite{Vaidman2020}, where the author concludes that \emph{additional assumptions are needed to derive the Born rule}.

\bibitem{Landsman2008} N. P. Landsman, \emph{The Born rule and its interpretation }, in  F. Weinert, K. Hentschel,  D. Greenberger, B. Falkenburg (eds.), Compendium of Quantum Physics (Springer 2008).
%doi: https://doi.org/10.1007/978-3-540-70626-7_20

\bibitem{Vaidman2020} L. Vaidman, \emph{Derivations of the Born Rule} in: Hemmo, M., Shenker, O. (eds) Quantum, Probability, Logic. Jerusalem Studies in Philosophy and History of Science (Springer, Cham, 2020).
%doi: https://doi.org/10.1007/978-3-030-34316-3_26

\bibitem{Dowling} J. P. Dowling and G. J. Milburn, \emph{Quantum technology: The second quantum revolution}, Philosophical Transactions of the Royal Society A, {\bf 361},  1655 (2003).
%doi: https://doi.org/10.1098/rsta.2003.1227

\bibitem{Deutsch2020} I. H. Deutsch, \emph{Harnessing the power of the second quantum revolution}, PRX Quantum 1, 020101 (2020).
%doi: https://doi.org/10.1103/PRXQuantum.1.020101

\bibitem{Haroche2013} S. Haroche, \emph{Nobel lecture: Controlling photons in a box and exploring the quantum to classical boundary}, Reviews of Modern Physics {\bf 85}, 1083 (2013).
%\doi: https://doi.org/10.1103/RevModPhys.85.1083 

\bibitem{Wineland2013} D. J. Wineland, \emph{Nobel lecture: Superposition, entanglement, and raising Schr\"{o}dinger's cat}, Reviews of Modern Physics {\bf 85}, 1103 (2013).
%doi: https://doi.org/10.1103/RevModPhys.85.1103

\bibitem{Wiseman} H. M. Wiseman, G. J. Milburn, \emph{Quantum Measurement and Control} (Cambridge University Press, 2009).

\bibitem{Peres1993} A. Peres, \emph{Quantum theory, concepts and methods}, (Springer, Dordrecht, 1993).
%doi: https://doi.org/10.1007/0-306-47120-5

\bibitem{Brun2002} T. A. Brun, \emph{A simple model of quantum trajectories}, American Journal of Physics {\bf 70}, 719 (2002).
%doi: https://doi.org/10.1119/1.1475328

\bibitem{Jacobs2006} K. Jacobs and D. A. Steck, \emph{A straightforward introduction to continuous quantum measurement}, Contemporary Physics {\bf 47}, 279 (2006).
%doi: https://doi.org/10.1080/00107510601101934

\bibitem{Tamir2013} B. Tamir and E. Cohen, \emph{Introduction to weak measurements and weak values}, Quanta {\bf 2}, 7 (2013).
%doi: http://dx.doi.org/10.12743/quanta.v2i1.14

\bibitem{Svensson2013} B. E. Y. Svensson, \emph{Pedagogical review of quantum measurement theory with an emphasis on weak measurements}, Quanta {\bf 2}, 18 (2013).
%doi: https://doi.org/10.12743/quanta.v2i1.12

\bibitem{Jacobs2014} K. Jacobs, \emph{Quantum measurement theory and its applications} (Cambridge University Press, Cambridge, UK, 2014). 
%

\bibitem{Dressel2014} J. Dressel, M. Malik, F. M. Miatto, A. N. Jordan, and R. W. Boyd, Colloquium: \emph{Understanding quantum weak values: Basics and applications}, Reviews of Modern Physics {\bf 86}, 307 (2014).
%doi: https://doi.org/10.1103/RevModPhys.86.307

\bibitem{Jordan2024} A. N. Jordan and I. A. Siddiqi, \emph{Quantum measurement: Theory and practice} (Cambridge University Press, Cambridge, UK, 2024).
% 

\bibitem{Gisin1984} N. Gisin, \emph{Quantum measurements and stochastic processes}, Physical Review Letters {\bf 52}, 1657 (1984).
%doi: https://doi.org/10.1103/PhysRevLett.52.1657

\bibitem{Patel2017} A. Patel and P. Kumar, \emph{Weak measurements, quantum-state collapse, and the Born rule}, Physical Review A {\bf 96}, 022108 (2017). 
%doi:https://doi.org/10.1103/PhysRevA.65.013813

\bibitem{Einstein1935} A. Einstein, B. Podolsky, and N. Rosen, \emph{Can quantum-mechanical description of physical reality be considered complete?}, Physical Review {\bf 47}, 777 (1935).
%doi: https://doi.org/10.1103/PhysRev.47.777

\bibitem{Bracken2021} C. Bracken, J. R. Hance, and S. Hossenfelder, \emph{The quantum eraser paradox}, arXiv:2111.09347.
%doi: https://arxiv.org/abs/2111.09347

\bibitem{Wheeler1978} J. A. Wheeler, in \emph{Mathematical Foundations of Quantum Theory}, ed. R. Marlow (Academic Press, New York, 1978).

\bibitem{Loudon} R. Loudon, \emph{The Quantum Theory of Light}, third edition, (Oxford University Press, 2000). 

\bibitem{Huang} K. Huang, \emph{Statistical mechanics},  2nd edition (Wiley, 1991).

\bibitem{Aspect1981} A. Aspect, P. Grangier, R. Gerard, \emph{Experimental tests of realistic local theories via Bell's theorem}, Physical Review Letters {\bf 47}, 460 (1981).
%doi:https://doi.org/10.1103/PhysRevLett.47.460

\bibitem{Aspect1982a} A. Aspect, P. Grangier, G. Roger, \emph{Experimental realization of Einstein-Podolsky-Rosen-Bohm Gedankenexperiment: A new violation of Bell's inequalities}, Physical Review Letters {\bf 49}, 91 (1982).
%doi:https://doi.org/10.1103/PhysRevLett.49.91

\bibitem{Aspect1982b} A. Aspect, J. Dalibard, G. Roger, \emph{Experimental Test of Bell's Inequalities Using Time-Varying Analyzers}, Physical Review Letters, {\bf 49}, 1804 (1982).
%doi:https://doi.org/10.1103/PhysRevLett.49.1804

\bibitem{Haroche2007} C. Guerlin, J. Bernu, S. Del\'eglise, C. Sayrin, S. Gleyzes, S. Kuhr, M. Brune, J.-M. Raimond, and S. Haroche, \emph{Progressive field-state collapse and quantum non-demolition photon counting}, Nature {\bf 448}, 889 (2007). 
%doi:https://doi.org/10.1038/nature06057

\bibitem{Zeilinger} M. Giustina, M. A. M. Versteegh, S. Wengerowsky, J. Handsteiner, A. Hochrainer, K. Phelan, F. Steinlechner, J. Kofler, J.-Å. Larsson, C. Abell{\'a}n, W. Amaya, V. Pruneri, M. W. Mitchell, J. Beyer, T. Gerrits, A. E. Lita, L. K. Shalm, S. W. Nam, T. Scheidl, R. Ursin, B. Wittmann, and A. Zeilinger, \emph{Significant-loophole-free test of Bell's theorem with entangled photons}, Physical Review Letters {\bf 115}, 250401 (2015).
%doi:http://dx.doi.org/10.1103/PhysRevLett.115.250401

\bibitem{Grangier} P. Grangier, G. Roger and A. Aspect, \emph{Experimental evidence for a photon anticorrelation effect on a beam splitter: A new light on single-photon interferences}, Europhysics Letters, {\bf 1}, 173 (1986). 
%doi:10.1209/0295-5075/1/4/004

\bibitem{Aspect2015} A. Aspect, \emph{Closing the door on Einstein and Bohr's quantum debate}, Physics {\bf 8}, 123 (2015).
%doi: https://doi.org/10.1103/Physics.8.123

\bibitem{Mott} N. F. Mott, \emph{The wave mechanics of $\alpha$-Ray tracks}, Proceedings of the Royal Society of London. Series A, {\bf 126}, 79  (1929).
%doi: https://doi.org/10.1098/rspa.1929.0205

\bibitem{Navez2023} P. Navez, A. G. Balanov, S. E. Savel'ev and A. M. Zagoskin, \emph{Quantum electrodynamics of non-demolition detection of single microwave photon by superconducting qubit array}, Journal Applied Physics {\bf 133}, 104401 (2023). 
%doi: https://doi.org/10.1063/5.0137747

\bibitem{Navez1997} P. Navez, D. Bitouk, M. Gajda, Z. Idziaszek, and K. Rzazewski, \emph{Fourth statistical ensemble for the Bose-Einstein condensate}, Physical Review Letters {\bf 79}, 1789  (1997).
%doi: https://doi.org/10.1103/PhysRevLett.79.1789

\bibitem{Ryder} Lewis H. Ryder, \emph{Quantum Field Theory} 2nd Edition, (Cambridge University Press, 1996).

\bibitem{Ouerdane2010} H. Ouerdane, B. Gervais, H. Zhou, M. Beuve, and J.-Ph. Renault, \emph{Radiolysis of water confined in porous silica: A simulation study of the physicochemical yields}, Journal of Physical Chemistry C {\bf 114}, 12667 (2010).
%doi: https://doi.org/10.1021/jp103127j

\bibitem{Martini}
E. Jeffrey, D. Sank, J. Y. Mutus, T. C. White, J. Kelly, R. Barends, Y. Chen, Z. Chen, B. Chiaro,
A. Dunsworth, A. Megrant, P. J. J. O’Malley, C. Neill, P. Roushan, A. Vainsencher, J. Wenner, A. N. Cleland, and J. M. Martinis, \emph{Fast accurate state measurement with superconducting qubits},
Physical Review Letters {\bf 112}, 190504 (2014).
%doi: https://doi.org/10.1103/PhysRevLett.112.190504

\bibitem{Braginsky1996} V. B. Braginsky and F. Ya. Khalili, \emph{Quantum nondemolition measurements: the route from toys to tools}, Reviews of Modern Physics {\bf 68}, 1 (1996).
%doi :https://doi.org/10.1103/RevModPhys.68.1

\bibitem{Blais2004a} A. Blais, R.-S. Huang, A. Wallraff, S. M. Girvin, and R. J. Schoelkopf, \emph{Cavity quantum electrodynamics for superconducting electrical circuits: An architecture for quantum computation} Physical Review A {\bf 69}, 062320 (2004).
%doi: https://doi.org/10.1103/PhysRevA.69.062320

\bibitem{Wallraff2024} A. Wallraff, D. I. Schuster, A. Blais, L. Frunzio, R.- S. Huang, J. Majer, S. Kumar, S. M. Girvin \& R. J. Schoelkopf, \emph{Strong coupling of a single photon to a superconducting qubit using circuit quantum electrodynamics}, Nature {\bf 431}, 162 (2004).
%doi: https://doi.org/10.1038/nature02851

\bibitem{Freedman1972} S. J. Freedman, and J. F. Clauser, \emph{Experimental test of local hidden-variable theories}, Physical Review Letters {\bf 28}, 938 (1972).
%doi: https://doi.org/10.1103/PhysRevLett.28.938

\bibitem{Weihs1998} G. Weihs, T. Jennewein, C. Simon, H. Weinfurter, and A. Zeilinger, \emph{Violation of Bell's inequality under strict Einstein locality conditions}, Physical Review Letters {\bf 81}, 5039 (1998).
%doi: https://doi.org/10.1103/PhysRevLett.81.5039

\bibitem{Rowe2001} M. A. Rowe, D. Kielpinski, V. Meyer, C. A. Sackett, W. M. Itano, C. Monroe and D. J. Wineland, \emph{Experimental violation of a Bell's inequality with efficient detection}, Nature {\bf 409}, 791 (2001).
%doi: https://doi.org/10.1038/35057215

\bibitem{Matsukevich2008} D. N. Matsukevich, P. Maunz, D. L. Moehring, S. Olmschenk, and C. Monroe, \emph{Bell inequality violation with two remote atomic qubits}, Physical Review Letters {\bf 100}, 150404 (2008).
%doi: https://doi.org/10.1103/PhysRevLett.100.150404

\bibitem{Ansmann2009} M. Ansmann, H. Wang, R. C. Bialczak, M. Hofheinz, E. Lucero, M. Neeley, A. D. O'Connell, D. Sank, M. Weides, J. Wenner, A. N. Cleland, and J. M. Martini, \emph{Violation of Bell's inequality in Josephson phase qubits}, Nature {\bf 461}, 504 (2009).
%doi: https://doi.org/10.1038/nature08363

\bibitem{Scheidl2010} T. Scheidl, R. Ursin, J. Kofler, S. Ramelow, X. Ma, T. Herbst, L. Ratschbacher, A. Fedrizzi, N.K. Langford, T. Jennewein, and A. Zeilinger, \emph{Violation of local realism with freedom of choice}, Proceedings of the National Academy of Sciences {\bf 107}, 19708 (2010).
%doi: https://doi.org/10.1073/pnas.1002780107

\bibitem{Giustina2013} M. Giustina, A. Mech, S. Ramelow, B. Wittmann, J. Kofler, J. Beyer, A. Lita, B. Calkins, T. Gerrits, S. W. Nam, R. Ursin, and A. Zeilinger \emph{Bell violation using entangled photons without the fair-sampling assumption}, Nature {\bf 497}, 227 (2013).
%doi: https://doi.org/10.1038/nature12012

\bibitem{Christensen2013} B. G. Christensen, K. T. McCusker, J. B. Altepeter, B. Calkins, T. Gerrits, A. E. Lita, A. Miller, L. K. Shalm, Y. Zhang, S. W. Nam, N. Brunner, C. C. W. Lim, N. Gisin, and P. G. Kwiat, \emph{Detection-loophole-free test of quantum nonlocality, and applications}, Physical Review Letters {\bf 111}, 130406 (2013).
%doi: https://doi.org/10.1103/PhysRevLett.111.130406

\bibitem{Handsteiner2017} J. Handsteiner, A. S. Friedman, D. Rauch, J. Gallicchio, B. Liu, H. Hosp, J. Kofler, D. Bricher, M. Fink, C. Leung, A. Mark, H. T. Nguyen, I. Sanders, F. Steinlechner, R. Ursin, S. Wengerowsky, A. H. Guth, D. I. Kaiser, T. Scheidl, and A. Zeilinger, \emph{Cosmic Bell test: Measurement settings from Milky Way stars}, Physical Review Letters {\bf 118}, 060401 (2017).
%doi: 

\bibitem{Bell1964} J. S. Bell, \emph{On the Einstein–Podolsky–Rosen paradox}, Physics {\bf 1}, 195 (1964).
%doi: https://doi.org/10.1103/PhysicsPhysiqueFizika.1.195

\bibitem{Brunner2014} N. Brunner, D. Cavalcanti, S. Pironio, V. Scarani, and S. Wehner, \emph{Bell nonlocality}, Reviews of Modern Physics {\bf 86}, 419 (2014).
%doi: https://doi.org/10.1103/RevModPhys.86.419

\bibitem{Hensen2015} B. Hensen,  H. Bernien, A. E. Dr\'eau, A. Reiserer, N. Kalb, M. S. Blok, J. Ruitenberg, R. F. L. Vermeulen, R. N. Schouten, C. Abell\'an, W. Amaya, V. Pruneri, M. W. Mitchell, M. Markham, D. J. Twitchen, D. Elkouss, S. Wehner, T. H. Taminiau, and R. Hanson, \emph{Loophole-free Bell inequality violation using electron spins separated by 1.3 kilometres}, Nature {\bf 526}, 682 (2015).
%doi: https://doi.org/10.1038/nature15759

\bibitem{Bell1987} J. S. Bell, {\it Speakable and Unspeakable in Quantum Mechanics} (Cambridge University Press, 1987). 
    
\bibitem{Kolobov1989} M. I. Kolobov and I. V. Sokolov, \emph{Spatial behavior of squeezed states of light and quantum noise in optical images}, Soviet Physics JETP {\bf 69}, 1097 (1989).
%link: http://jetp.ras.ru/cgi-bin/dn/e_069_06_1097.pdf

\bibitem{Navez2001} P. Navez, E. Brambilla, A. Gatti, and L. A. Lugiato, \emph{Spatial entanglement of twin quantum images}, Physical Review A {\bf 65}, 013813 (2001).

\bibitem{Bricmont} J. Bricmont, S. Goldstein, and D. Hemmick, \emph{From EPR-Schr\"{o}dinger paradox to nonlocality based on perfect correlations}, Foundations of Physics {\bf 52}, 53 (2022).
%doi: https://doi.org/10.1007/s10701-022-00568-8

\end{thebibliography}
\end{document}